\documentclass[aps,prd,preprint,superscriptaddress,nofootinbib]{revtex4-1}
\usepackage{braket}
\usepackage{xargs}
\usepackage{mathtools}
\usepackage[english]{babel}
\usepackage[utf8]{inputenc}
\usepackage{amsmath,amssymb,braket,slashed,mathrsfs}
\usepackage{pifont}
\usepackage{graphicx}
\usepackage{color,hyperref}
\usepackage{multirow,array}

\newcommand{\ba}{\begin{eqnarray}}
\newcommand{\ea}{\end{eqnarray}}
\newcommand{\be}{\begin{equation}}
\newcommand{\ee}{\end{equation}}
\newcommand{\nn}{\nonumber}

\newcommand{\innovation}{Collaborative Innovation Center of Quantum Matter, Beijing 100871, China}
\newcommand{\chep}{Center for High Energy Physics, Peking University, Beijing 100871, China}
\newcommand{\pkuphy}{School of Physics, Peking University, Beijing 100871,
China}

\newcommand{\Uconn}{Department of Physics, University of Connecticut, Storrs, CT 06269, USA}
\newcommand{\RBRC}{RIKEN-BNL Research Center, Brookhaven National Laboratory, Building 510, Upton, NY 11973}

\begin{document}

	\title{Lattice QCD calculation of light sterile neutrino contribution \\in $0\nu2\beta$ decay}
	
	\author{Xin-Yu Tuo}\affiliation{\pkuphy}
	\author{Xu~Feng}\email{xu.feng@pku.edu.cn}\affiliation{\pkuphy}\affiliation{\innovation}\affiliation{\chep}
	\author{Lu-Chang Jin}\email{ljin.luchang@gmail.com}\affiliation{\Uconn}\affiliation{\RBRC}

	\date{\today}
	
	\begin{abstract}
	
	We present a lattice QCD study of the neutrinoless double beta decay involving light sterile neutrinos.
	The calculation is performed at physical pion mass using five gauge ensembles generated with $2+1$-flavor domain wall fermions. 
	We obtain the low-energy constants $g_{\text{LR}}^{\pi\pi}(m_\nu)$ with the neutrino mass $m_\nu$ from $0$ GeV to $3$ GeV. 
	The lattice results are reasonably consistent with the previous interpolation method~\cite{Dekens:2020ttz} with a $\sim20$\% deviation at small $m_\nu$.
	We provide an explanation on the discrepancy at vanishing neutrino mass.
	At large $m_\nu$, a good consistency between our results and the previous lattice determination of $g_{4}^{\pi\pi}(\mu)$~\cite{Nicholson:2018mwc} is found at $\mu=m_\nu=3$ GeV.
	\end{abstract}
	
	\maketitle

\section{Introduction}
\label{Intro}

Observation of new physics beyond the Standard Model (BSM) is one of the most central themes in particle physics. 
It is a milestone that the neutrino oscillations are observed and the neutrinos are demonstrated as massive particles. To further determine the absolute mass of the neutrinos and verify whether the neutrinos are Dirac or Majorana fermions, neutrinoless double beta ($0\nu2\beta$) decay experiments~\cite{Agostini:2019hzm,KamLAND-Zen:2016pfg,Alduino:2017ehq,Albert:2017owj,Azzolini:2018dyb,Alvis:2019sil,GERDA:2020xhi,CUPID:2020aow,CUORE:2021mvw,KamLAND-Zen:2022tow} provide an excellent probe.
In the framework of effective field theory (EFT), a roadmap to reach the theoretical determination of the $0\nu2\beta$ decay amplitude is laid out through a path from the BSM scenarios at high energy towards the nuclear many-body theories at low energy~\cite{Cirigliano:2022oqy}. At the energy of 
$O(1)$ GeV - $O(100)$ MeV, chiral perturbation theory ($\chi$PT) and its extension to the multi-nucleon sector, namely chiral EFT, serve as a bridge to connect quark-level theories and nuclear-level theories~\cite{Cirigliano:2017ymo,Cirigliano:2017djv,Cirigliano:2017tvr,Pastore:2017ofx,Cirigliano:2018hja,Cirigliano:2018yza,Cirigliano:2019vdj,Dekens:2020ttz,Cirigliano:2020dmx,Cirigliano:2021qko}. To construct the hadronic operators, it requires  the non-perturbative inputs  of the low-energy couplings from lattice QCD~\cite{Cirigliano:2020yhp}.

In recent years lattice QCD calculations have successfully provided the hadronic matrix elements for double neutrino double beta decay in the nucleon sector~\cite{Shanahan:2017bgi,Tiburzi:2017iux} and $0\nu2\beta$ matrix elements in the pion sector including both long-distance contributions via Majorana neutrino exchange~\cite{Feng:2018pdq,Tuo:2019bue,Detmold:2020jqv} and short-distance contributions via dim-9 operators~\cite{Nicholson:2018mwc}. The approaches to match the lattice matrix elements from finite volume to the double beta decay amplitudes in the infinite volume are also proposed~\cite{Feng:2020nqj,Davoudi:2020xdv,Davoudi:2020gxs,Davoudi:2021noh}.

Among various BSM models, the inclusion of additional sterile neutrinos presents an important class of the extension of the Standard Model, as it can be used to explain
the matter/antimatter asymmetry via low-scale leptogenesis~\cite{Akhmedov:1998qx,Asaka:2005an,Asaka:2005pn,Shaposhnikov:2006nn,Canetti:2012vf,Hernandez:2016kel,Ghiglieri:2017gjz}.
If these neutrinos are Majorana particles, they can mediate and influence the $0\nu 2\beta$ decay rates. Understanding the impact of massive sterile neutrinos can help $0\nu2\beta$ experiments provide more constraints on neutrino mass scenarios. 
As such, the EFT description of the $0\nu 2\beta$ decay has been extended to include the option of light sterile neutrinos~\cite{Dekens:2020ttz}. Using an effective interpolation method it is found that 
the decay amplitudes peak at the neutrino mass in a range from few hundreds MeV to few GeV. The exact location of peak would rely on the input of the underlying operators. 
The determination of them is highly non-trivial due to non-perturbative nature of QCD.
As a consequence, it brings in sizable uncertainties at the order-of-magnitude level, which are dominated by the  low-energy constants (LECs). It motivates our lattice QCD study of the 
non-trivial neutrino mass dependence of LECs, which is reported here.

\section{Theoretical background}

Here we follow the systematic EFT approach~\cite{Dekens:2020ttz} to add the massive sterile neutrinos in $0\nu2\beta$ decay.
At the energy above the electroweak scale $\Lambda_{\text{EW}}$, the SM is extended with sterile neutrinos in the framework of SMEFT~\cite{delAguila:2008ir,Cirigliano:2012ab,Asaka:2016zib,Liao:2016qyd,Liao:2019tep,Liao:2020zyx,Liao:2020jmn} and the higher-dimensional operators are included when the non-neutrino states heavier than $\Lambda_{\text{EW}}$ are integrated out. To describe physics at or below 
$\Lambda_{\text{EW}}$, it proceeds in two different ways depending on the size of the neutrino mass $m_\nu$. For $\Lambda_\chi<m_\nu<\Lambda_{\text{EW}}$ where $\Lambda_\chi$ is chiral symmetry breaking scale, the sterile neutrinos are integrated out and effective local operators are produced.
For $m_\nu < \Lambda_\chi$, the neutrino remains an active degree of freedom propagating at the hadronic scale. In this case, one needs to obtain the $0\nu2\beta$ transition operators induced by light sterile neutrino in the framework of chiral EFT. The relevant nuclear matrix elements (NMEs) and the LECs in the chiral Lagrangian are all non-perturbative, and thus their dependence on the mass of the sterile neutrinos is poorly known. In Ref.~\cite{Dekens:2020ttz} naive interpolation formulae grounded in QCD and $\chi$PT are proposed to connect the regimes with $m_\nu \ll \Lambda_\chi$ and 
$m_\nu \gg \Lambda_\chi$.  Further improvements can be made if the relevant pion, nucleon and two-nucleon matrix elements are calculated and the LECs are extracted using non-perturbative techniques such as lattice QCD.
As the last step, the NMEs and the LECs are used to estimate the $0\nu2\beta$ half-lives. Both the uncertainties on the NMEs and LECs play a significant role in the determination of the $0\nu2\beta$ half-lives and hence the comparison with the experiments.

As mentioned above, lattice QCD can provide useful LECs for the construction of hadronic operators defined in $\chi$PT and chiral EFT. For simplicity, at low energy we discuss the scenario involving one neutrino mass eigenstate with mass $m_\nu$ which couples to left-handed electrons and to both right- and left-handed quark vector currents. Namely, the dim-6 operators describing the four-fermion interaction are introduced as
\be
\label{eq:dim-6}
\mathcal{L}^{(6)}=\frac{2 G_{F}}{\sqrt{2}}\left[C_{\mathrm{VLL}}^{(6)}\bar{u}_{L} \gamma^{\mu} d_{L}\bar{e}_{L} \gamma_{\mu} \nu+C_{\mathrm{VRL}}^{(6)}\bar{u}_{R} \gamma^{\mu} d_{R}\bar{e}_{L} \gamma_{\mu} \nu\right]+\text{h.c.},
\ee
where $G_F$ is Fermi constant, $u,d$ are quark fields and $e,\nu$ are electron and neutrino fields. Although only one neutrino mass eigenstate is involved here, the operator can be easily extended to include more mass eigenstates. Here $C_{\mathrm{VLL}}^{(6)}$ and $C_{\mathrm{VRL}}^{(6)}$ are Wilson coefficients which factorize information of underlying short-distance contribution of the BSM scenarios. 
Note that we do not include the scalar and tensor operators as well as higher dimensional operators. Thus, up to the Wilson coefficients,
one can view Eq.~(\ref{eq:dim-6}) as a simple extension of Standard Model to include the coupling of neutrino to right-handed quark current. 

The $0\nu2\beta$ decay involves the double current insertions with the vector couplings $C_{\mathrm{VLL,VRL}}^{(6)}$. The general amplitude in the hadronic sector is given by
\be
\label{eq:transition_amplitude}
\left\langle h_f e_1e_2\left|\frac{i}{2!}\int d^4x \,T\left\{\mathcal{L}^{(6)}(x)\mathcal{L}^{(6)}(0)\right\}\right|h_i\right\rangle,
\ee
where $h_{i,f}$ are hadronic states. The decay amplitude at the quark level shall be matched to the one at the hadronic level using chiral Lagrangian. At leading order, both the Lagrangian $\mathcal{L}_{\pi\pi}$ describing the $\pi^-\to\pi^+ ee$ transition and $\mathcal{L}_{NN}$ appearing in the nucleon-nucleon sector are important. Eq.~(\ref{eq:transition_amplitude}) does not induce $\pi N$ couplings at leading order and thus we neglect them. The pionic Lagrangian is given by
\be
\mathcal{L}_{\pi \pi}=2 G_{F}^{2} F_{\pi}^{2}C_{\mathrm{VLL}}^{(6)}C_{\mathrm{VRL}}^{(6)} m_{\nu} g_{\mathrm{LR}}^{\pi \pi}\left(m_{\nu}\right) \mathrm{Tr}\left[\mathcal{Q}_{L} \mathcal{Q}_{R}\right] \bar{e}_{L} e_{L}^{c},
\ee
with $\mathcal{Q}_{L}=u^\dagger \tau^+ u$, $\mathcal{Q}_{R}=u\tau^+u^\dagger$ and $\tau^\pm=(\tau_1\pm i\tau_2)/2$, where $\tau_i$ are the Pauli matrices and $u$ is a $2\times2$ matrix containing the pion fields and transforming as $u\to LuK^\dagger=KuR^\dagger$ under $SU(2)_L\times SU(2)_R$ transformations. Here $F_\pi$ is a pion decay constant. 
The terms associated with $(C_{\mathrm{VLL}}^{(6)})^2$ and $(C_{\mathrm{VRL}}^{(6)})^2$ produce the LEC $g_\nu^{\pi\pi}(m_\nu)$, but are considered as higher-order contributions and thus neglected.
 The remaining 
$C_{\mathrm{VLL}}^{(6)}C_{\mathrm{VRL}}^{(6)}$ term contains the LEC $g_{\mathrm{LR}}^{\pi \pi}(m_\nu)$, which is the main objective of this work. For the nucleonic Lagrangian $\mathcal{L}_{NN}$ which contains the LECs $g_{\mathrm{LR}}^{NN}(m_\nu)$ and $g_{\mathrm{\nu}}^{NN}(m_\nu)$, we leave it for the future study. 

Considering the similarity between the exchange of the neutrinos in $0\nu2\beta$ decay and the exchange of photons in the electromagnetic correction to the pion mass,
$g_{\mathrm{LR}}^{\pi \pi}(m_\nu)$ at $m_\nu=0$ can be approximated by~\cite{Dekens:2020ttz}
\be
\label{eq:mass_splitting}
g_{\text{LR}}^{\pi\pi}(m_\nu=0)\approx\frac{m_{\pi^\pm}^2-m_{\pi_0}^2}{2e^2}\approx 0.8F_\pi^2.
\ee
On the other hand, when the neutrino mass significantly exceed the scale $\Lambda_\chi$, the operator product of $\mathcal{L}^{(6)}$ in Eq.~(\ref{eq:transition_amplitude}) will induce dimension-9 operators involving four quark and two lepton fields, namely 
$O_4=\bar{q}^\alpha_L\gamma_\mu \tau^+ q_L^\alpha\,\bar{q}_R^\beta\gamma^\mu\tau^+q^\beta_R$ and its color-mixing pattern $O_5=\bar{q}^\alpha_L\gamma_\mu \tau^+ q_L^\beta\,\bar{q}_R^\beta\gamma^\mu\tau^+q^\alpha_R$ with $\alpha$, $\beta$ color indices. The pion matrix elements involving $O_{4,5}$ can be used to extract the additional LECs $g_{4,5}^{\pi\pi}(\mu)$ through a leading-order $\chi$PT formula 
\be
g_{4,5}^{\pi\pi}(\mu)=\frac{1}{F_\pi^2}\langle \pi|O_{4,5}(\mu)|\pi\rangle.
\ee
Thus one can match $g_{\text{LR}}^{\pi\pi}(m_\nu)$ at $m_\nu\gg \Lambda_\chi$ to $g_{4,5}^{\pi\pi}(\mu)$ as
\be
g_{\text{LR}}^{\pi\pi}(m_\nu)=C_4(m_\nu,\mu)g_4^{\pi\pi}(\mu)+C_5(m_\nu,\mu)g_5^{\pi\pi}(\mu),
\ee
where $C_{4,5}$ are the matching coefficients. Setting $\mu=m_\nu$, one has $C_4(m_\nu,\mu)=-\frac{F_\pi^2}{4m_\nu^2}+\mathcal{O}(\alpha_s)$ and $C_5(m_\nu,\mu)=\mathcal{O}(\alpha_s)$. The renormalization group equation (RGE) of $g_{4,5}^{\pi\pi}$ yields~\cite{Dekens:2020ttz}
\begin{equation}
\label{run2}
	\frac{d}{d \ln \mu}\left(\begin{array}{l}
	g_{4}^{\pi \pi} \\
	g_{5}^{\pi \pi}
	\end{array}\right)=-\frac{\alpha_{s}}{4 \pi}\left(\begin{array}{cc}
	6 / N_{c} & 0 \\
	-6 & -12 C_{F}
	\end{array}\right)^{T}\left(\begin{array}{l}
	g_{4}^{\pi \pi} \\
	g_{5}^{\pi \pi}
	\end{array}\right).
\end{equation}
Using the direct lattice QCD calculation of $\langle \pi|O_{4,5}(\mu)|\pi\rangle$~\cite{Nicholson:2018mwc} at $\mu=3$ GeV as input, one obtains
$g_{4}^{\pi \pi}(\mu)$ at $\mu=2$ GeV~\cite{Dekens:2020ttz} 
\begin{equation}\label{run3}
	g_{4}^{\pi \pi}(\text{2 GeV})=-1.9(2) \text{ GeV}^{2},\quad g_{5}^{\pi \pi}(\text{2 GeV})=-8.0(6) \text{ GeV}^{2}.
\end{equation}

In this work we plan to perform a direct lattice QCD calculation of $g_{\text{LR}}^{\pi\pi}(m_\nu)$ in the range of 0 GeV $\le m_\nu\le 3$ GeV. The choice of $m_\nu$ allows us to perform a consistency check of Eq.~(\ref{eq:mass_splitting}) at $m_\nu=0$ and the previous lattice results of $g_{4}^{\pi \pi}(\mu)$ at $\mu=3$ GeV~\cite{Nicholson:2018mwc} by assuming 
$g_{\text{LR}}^{\pi\pi}(m_\nu)\simeq -\frac{F_\pi^2}{4m_\nu^2}g_4^{\pi\pi}(m_\nu)$. In addition, the lattice results of $g_{\text{LR}}^{\pi\pi}(m_\nu)$ can be used to produce the $0\nu2\beta$ amplitude through
\begin{eqnarray}
\label{AL}
\mathcal{A}_L=-\frac{m_\nu}{4m_e}&&\left\{2C_{\mathrm{VLL}}^{(6)}C_{\mathrm{VRL}}^{(6)} \left(\overline{\mathcal{M}}_V(m_\nu)-\overline{\mathcal{M}}_A(m_\nu)\right)\right.\nonumber\\&&\left.+\left[\left(C_{\mathrm{VLL}}^{(6)}\right)^2+\left(C_{\mathrm{VRL}}^{(6)}\right)^2\right]\left(\overline{\mathcal{M}}_V(m_\nu)+\overline{\mathcal{M}}_A(m_\nu)\right) \right\},      
\end{eqnarray}
where
\begin{subequations}
	\begin{equation}
	\overline{\mathcal{M}}_V(m_\nu)+\overline{\mathcal{M}}_A(m_\nu)=\mathcal{M}_V(m_\nu)+\mathcal{M}_A(m_\nu)+\frac{2m_\pi^2}{g_A^2}g_\nu^{NN}(m_\nu)\mathcal{M}_{F,sd}
	\end{equation}
	\begin{equation}
	\overline{\mathcal{M}}_V(m_\nu)-\overline{\mathcal{M}}_A(m_\nu)=\mathcal{M}_V(m_\nu)-\mathcal{M}_A(m_\nu)+\frac{2m_\pi^2}{g_A^2}g_{\text{LR}}^{NN}(m_\nu)M_{F,sd}-4g_{\text{LR}}^{\pi\pi}(m_\nu)\mathcal{M}_{PS,sd}.
	\end{equation}
\end{subequations}
Here long-distance NMEs $\mathcal{M}_V(m_\nu)$ and $\mathcal{M}_A(m_\nu)$ and short-distance ones $\mathcal{M}_{F,sd}$ and $\mathcal{M}_{PS,sd}$ 
are defined in Ref.~\cite{Dekens:2020ttz}. Various nuclear many-body methods such as quasi-particle random phase approximation (QRPA)~\cite{Jokiniemi:2018sxd} and the Shell Model~\cite{Menendez:2017fdf} are used
to calculate these NMEs for $^{76}$Ge, $^{82}$Se, $^{130}$Te and $^{136}$Xe and the results are listed in Ref.~\cite{Dekens:2020ttz}. Unfortunately, the LECs $g_\nu^{NN}(m_\nu)$ and $g_{\text{LR}}^{NN}(m_\nu)$ are poorly known.
In Ref.~\cite{Dekens:2020ttz} the uncertainties from these LECs are included by assigning a 50\% error on the contribution from $g_{\text{LR}}^{\pi\pi}(m_\nu)$ under the assumption that $g_\nu^{NN}(m_\nu)$ and $g_{\text{LR}}^{NN}(m_\nu)$ are similar as $g_{\text{LR}}^{\pi\pi}(m_\nu)$ in the order of magnitude. In this work, we simply ignore them. As it can be found later, the contribution from $g_{\text{LR}}^{\pi\pi}(m_\nu)$ produces a peak shape in the mass dependence of decay amplitude. 
Such shape is easy to understand as in the limit of $m_\nu\to0$ the decay amplitude is proportional to $m_\nu$, while in the limit of $m_\nu\to\infty$ the 
amplitude is suppressed by another factor of $1/m_\nu^2$ from the neutrino propagator. Thus the amplitude peaks at the intermediate neutrino mass.

\section{Lattice calculation}

To determine the LECs $g_{\text{LR}}^{\pi\pi}(m_\nu)$, we start with $\pi^-\to\pi^+ee$ transition amplitude defined using Eq.~(\ref{eq:transition_amplitude}), namely
\ba
\label{eq:amplitude1}
\mathcal{A}(\pi^-\to\pi^+ee)&=&\left\langle \pi^+ e_1e_2\left|\frac{i}{2!}\int d^4x \,T\left\{\mathcal{L}^{(6)}(x)\mathcal{L}^{(6)}(0)\right\}\right|\pi^-\right\rangle
\nn\\
&=&i\,G_F^2C_{\text{VLL}}^{(6)}C_{\text{VRL}}^{(6)}\langle e_1 e_2|\bar{e}_Le_L^c|0\rangle
\nn\\
&&\quad\times \int d^4x\,\langle \pi^+|T\left\{\bar{u}_L\gamma^\mu d_L(x)\bar{u}_R\gamma_\mu d_R(0)\right\}|\pi^-\rangle \,m_\nu S_0(x).
\ea
Here the term $m_\nu S_0(x)$ arises from the neutrino exchange with $S_0(x)$ the scalar propagator.
As sandwiched by left-handed electron field $\bar{e}_L$ and its charge conjugate $e_L^c$, only the mass term in the neutrino propagator can survive.

One then splits $\mathcal{A}(\pi^-\to\pi^+ee)$ into two parts
\be
\mathcal{A}(\pi^-\to\pi^+ee)=\mathcal{A}^{(0)}(\pi^-\to\pi^+ee)+\mathcal{A}^{(1)}(\pi^-\to\pi^+ee),
\ee
with $\mathcal{A}^{(0)}(\pi^-\to\pi^+ee)$ defined as
\ba
\mathcal{A}^{(0)}(\pi^-\to\pi^+ee)&=&i\,G_F^2C_{\text{VLL}}^{(6)}C_{\text{VRL}}^{(6)}\langle e_1 e_2|\bar{e}_Le_L^c|0\rangle
\nn\\
&&\quad\times\langle \pi^+|\bar{u}_L\gamma^\mu d_L(0)|0\rangle\langle0|\bar{u}_R\gamma_\mu d_R(0)|\pi^-\rangle \,m_\nu \tilde{S}_0(0).
\ea
Here $\tilde{S}_0(q)$ the Fourier transformation of propagator $S_0(x)$ with a factor of $e^{iqx}$. Note that in $\mathcal{A}^{(0)}(\pi^-\to\pi^+ee)$ the neutrino is assigned with vanishing momentum.  There is no loop integral resulting in the short-distance dim-9 operators. Thus only $\mathcal{A}^{(1)}(\pi^-\to\pi^+ee)$ is related to the chiral Lagrangian $\mathcal{L}_{\pi\pi}$. We have
\be
\label{eq:amplitude2}
\mathcal{A}^{(1)}(\pi^-\to\pi^+ee)=\langle \pi^+e_1e_2|\mathcal{L}_{\pi\pi}|\pi^-\rangle 
=4G_F^2C_{\text{VLL}}^{(6)}C_{\text{VRL}}^{(6)}\langle e_1e_2|\bar{e}_Le_L^c|0\rangle m_\nu g_{\text{LR}}^{\pi\pi}(m_\nu).
\ee
Combining Eqs.~(\ref{eq:amplitude1}) and (\ref{eq:amplitude2}) yields the matching condition
\ba
g_{\text{LR}}^{\pi\pi}(m_\nu)&=&\frac{i}{4}\int d^4x\,\langle \pi^+|T\{\bar{u}_L\gamma^\mu d_L(x)\bar{u}_R\gamma_\mu d_R(0)\}|\pi^-\rangle \, S_0(x)
\nn\\
&&-\frac{i}{4}\langle \pi^+|\bar{u}_L\gamma^\mu d_L(0)|0\rangle\langle0|\bar{u}_R\gamma_\mu d_R(0)|\pi^-\rangle \, \tilde{S}_0(0).
\ea

Using lattice QCD we calculate $g_{\text{LR}}^{\pi\pi}(m_\nu)$ in the Euclidean spacetime through the relation
\be
\label{eq:integral}
g_{\text{LR}}^{\pi\pi}(m_\nu)=-\frac{1}{16}\int d^4x \,S_0^E(x)\left[H_{VV}^E(x)-H_{AA}^E(x)\right],
\ee
where the pion matrix elements $H_{VV}^E(x)$ and $H_{AA}^E(x)$ are defined as
\ba
&&H_{VV}^E(x)=\langle \pi^+|T\left\{\bar{u}\gamma_\mu d(x)\bar{u}\gamma_\mu d(0)\right\}|\pi^-\rangle,
\nn\\
&&H_{AA}^E(x)=\langle \pi^+|T\left\{\bar{u}\gamma_\mu\gamma_5 d(x)\bar{u}\gamma_\mu\gamma_5 d(0)\right\}|\pi^-\rangle-H_0^E(x),
\nn\\
&&H_0^E(x)=\langle \pi^+|\bar{u}\gamma_\mu\gamma_5 d(x)|0\rangle\langle0|\bar{u}\gamma_\mu\gamma_5 d(0)|\pi^-\rangle
+\langle \pi^+|\bar{u}\gamma_\mu\gamma_5 d(0)|0\rangle\langle0|\bar{u}\gamma_\mu\gamma_5 d(x)|\pi^-\rangle.
\nn\\
\ea
Here $\gamma_\mu$ and $\gamma_5$ are the Euclidean gamma matrices. 
The scalar propagator 
$S_0^E(x)$ is related to the modified Bessel function of the second kind $K_1(\rho)$ through
\be
S_0^E(x)=\int\frac{d^4q}{(2\pi)^4}\frac{e^{-iqx}}{q^2+m_\nu^2}=\frac{m_\nu}{4\pi^2|x|}K_1(m_\nu|x|).
\ee

In Eq.~(\ref{eq:integral}) when the two currents approach each other one needs to examine the ultraviolet behavior of bilocal matrix elements.
Generally speaking, once the ultraviolet divergences appear, the renormalization procedure is required to remove the unphysical lattice cutoff effects. It has been discussed
in details~\cite{Christ:2016eae,Bai:2017fkh} how to make the renormalization in the RI-MOM scheme for the bilocal operators with two current insertions. Fortunately, the situation for the case of $g_{\text{LR}}^{\pi\pi}(m_\nu)$  is much simpler.  At the hard momentum region with 
$\Lambda\gg m_\nu,\Lambda_{\text{QCD}}$,
the operator product expansion in the continuum theory yields
\be
\int d^4x\,e^{i\Lambda x}\mathcal{L}^{(6)}(x)\mathcal{L}^{(6)}(0)\sim G_F^2C_{\text{VLL}}^{(6)}C_{\text{VRL}}^{(6)} \frac{m_\nu}{\Lambda^2} \mathcal{O}^{(9)},
\ee
where  the dim-9 operator $\mathcal{O}^{(9)}$ is consisted of four quark and two electron fields. By power counting, the ultraviolet contributions are suppressed by a factor of $1/\Lambda^2$. Within lattice QCD, the short-distance region is cutoff by the lattice spacing $a$ if $m_\nu\ll \frac{\pi}{a}$. Thus, the unphysical short-distance part only captures $\mathcal{O}(a^2)$ discretization effects in the calculation of $g_{\text{LR}}^{\pi\pi}(m_\nu)$.

\subsection{\label{sec:SE}Lattice setup}
In practice the hadronic functions $H^E_{VV}(x)$ and $H^{E}_{AA}(x)$ are calculated by constructing the four-point correlation function of
\ba
&&C_{JJ}(t_f,x,y,t_i)=\langle \phi_{\pi^+}(t_f) J(x) J(y)\phi^\dagger_{\pi^-}(t_i)\rangle,
\nn\\
&&C_{0}(t_f,x,y,t_i)=\langle \phi_{\pi^+}(t_f) A(x)\rangle\langle A(y)\phi^\dagger_{\pi^-}(t_i)\rangle
 +\langle \phi_{\pi^+}(t_f) A(y)\rangle\langle A(x)\phi^\dagger_{\pi^-}(t_i)\rangle,
\ea
where $J$ can be either charged vector ($V$) or axial-vector current $(A)$.
$\phi_{\pi^+}$ and $\phi^\dagger_{\pi^-}$ are Coulomb gauge-fixed wall-source operators at $t_f=\max(t_x,t_y)+\Delta T$ and $t_i=\min(t_x,t_y)-\Delta T$. $\Delta T$ is chosen to be sufficiently large for pion state dominance. For more details of lattice calculation of this four-point function, we refer readers to our previous calculation of $g_\nu^{\pi\pi}$ in Ref.~\cite{Tuo:2019bue}.

By constructing the ratio between four-point and two-point correlation functions, hadronic functions can be calculated via
\ba
&&H_{VV}^{(\text{lat})}(x-y)=\frac{C_{VV}(t_f,x,y,t_i)}{C_\pi(t_f,t_i)},\quad C_\pi(t_f,t_i)=\frac{1}{L^3}\langle \phi_{\pi^+}(t_f) \phi^\dagger_{\pi^+}(t_i)\rangle
\nn\\
&&H_{AA}^{(\text{lat})}(x-y)=\frac{C_{AA}(t_f,x,y,t_i)-C_0(t_f,x,y,t_i)}{C_\pi(t_f,t_i)}
\ea
where $L$ is the spatial extent of the lattice. The superscript (lat) is used to remind us that these quantities contain various systematic uncertainties such as finite-volume effects and lattice artifacts and thus differ from $H_{JJ}^{E}(x)$ used in Eq.~(\ref{eq:integral}).

In our calculation, the neutrino mass ranges from 0 GeV to 3 GeV. At the limit of $m_\nu\to0$, the neutrino is very soft and thus causes the similar finite-volume effects as that induced by the photon in the QCD+QED simulation. We adopt the infinite-volume reconstruction method~\cite{Feng:2018qpx}, which is originally proposed to compute the electromagnetic corrections to the hadron mass splitting with only exponentially suppressed finite-volume effects. Upon its development, this method has been successfully applied to the calculation of pion mass splitting~\cite{Feng:2021zek} and extended to various electroweak processes involving photon or massless leptonic propagators~\cite{Christ:2020jlp,Christ:2020hwe,Tuo:2021ewr,Meng:2021ecs,Fu:2022fgh}. After applying this method, we find that 
the residual exponentially suppressed finite-volume effects, named as $\delta_{\text{IVR}}(L)$ by Ref.~\cite{Tuo:2019bue}, are negligible compared to the statistical errors.

As shown in the Table~\ref{ens}, we use five lattice gauge ensembles at the physical pion mass, generated by RBC
and UKQCD Collaborations using $2+1$-flavor domain
wall fermion~\cite{Blum:2014tka}. Ensembles 48I and 64I use the Iwasaki gauge action
in the simulation while the other three use Iwasaki+DSDR action. We use local vector and axial-vector
currents in the calculation. These currents are matched to the
conserved ones by multiplying the renormalization factors $Z_{V/A}$, whose values are quoted from Ref.~\cite{RBC:2014ntl}.

\begin{table}
	\begin{ruledtabular}
		\begin{tabular}{cccccccc}
			Ensemble&Gauge action&$m_\pi$[MeV]&$L^3\times T$&$a^{-1}$[GeV]& $N_{conf}$&$m_\pi L$&$\Delta T/a$\\
			\hline
			24D&Iwasaki+DSDR&142& $24^3\times 64$ &1.015& 91&3.3& 8\\
			\hline
			32D&Iwasaki+DSDR&142& $32^3\times 64$ &1.015& 56&4.5& 8\\
			\hline
			32Df&Iwasaki+DSDR&143& $32^3\times 64$ &1.378& 24&3.3& 10\\
			\hline
			48I&Iwasaki&135& $48^3\times 96$ &1.730& 33&3.8& 12\\
			\hline
			64I&Iwasaki&135& $64^3\times 128$ &2.359& 67&3.7& 18
		\end{tabular}
	\end{ruledtabular}
	\caption{\label{ens}Lattice ensembles used in this work.}
\end{table}

\subsection{Numerical results}
\begin{figure}
	\centering
	\includegraphics[width=0.8\textwidth]{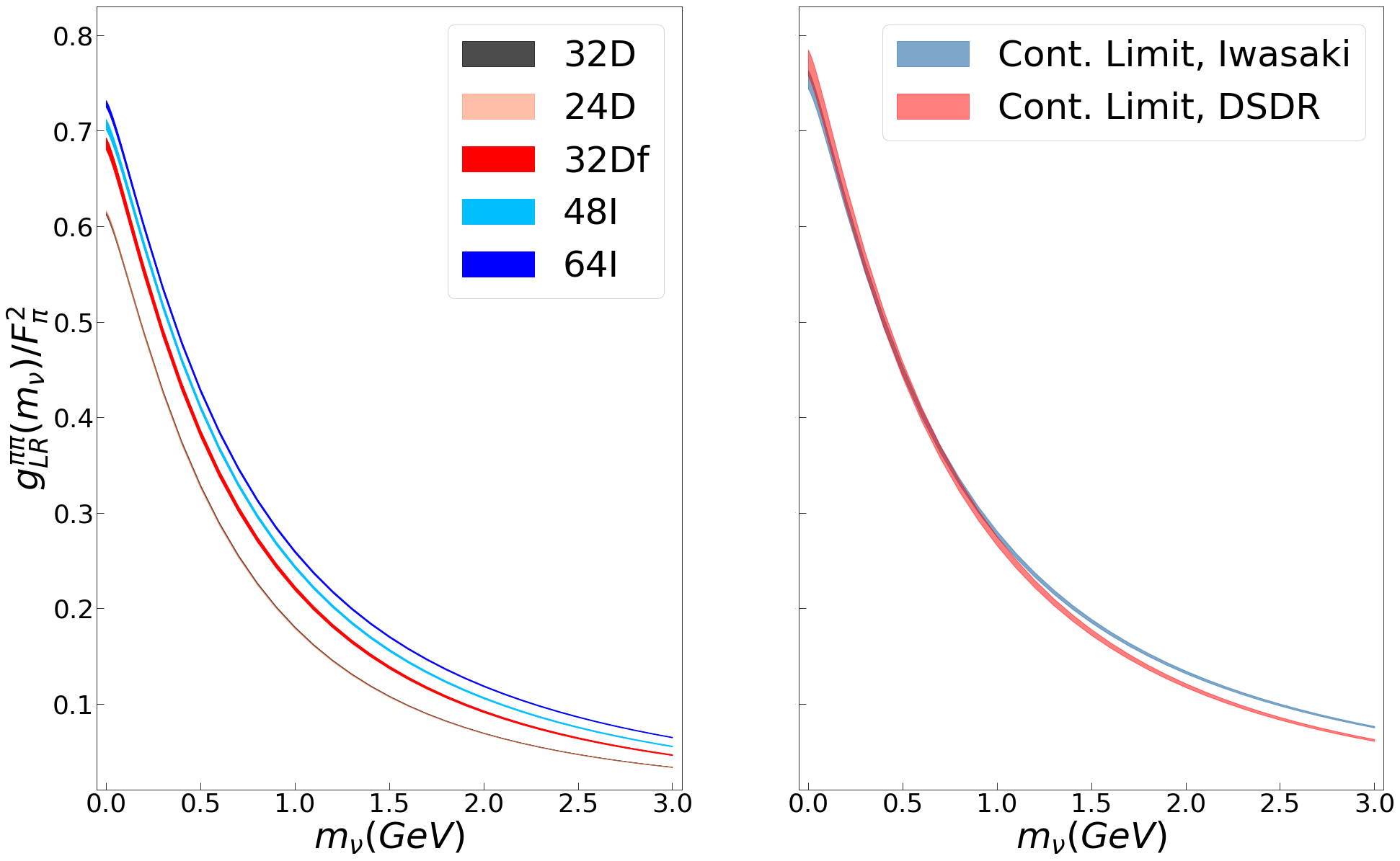}
	\caption{\label{fig:extra}Lattice results of $g_{\text{LR}}^{\pi\pi}(m_\nu)$ for each gauge ensemble (Left) and the continuum extrapolation based on Iwasaki and DSDR ensembles (Right). The 24D results are completely consistent with 32D results, making the latter nearly invisible in the figure.}
\end{figure}

\begin{figure}
	\centering
	\includegraphics[width=1.0\textwidth]{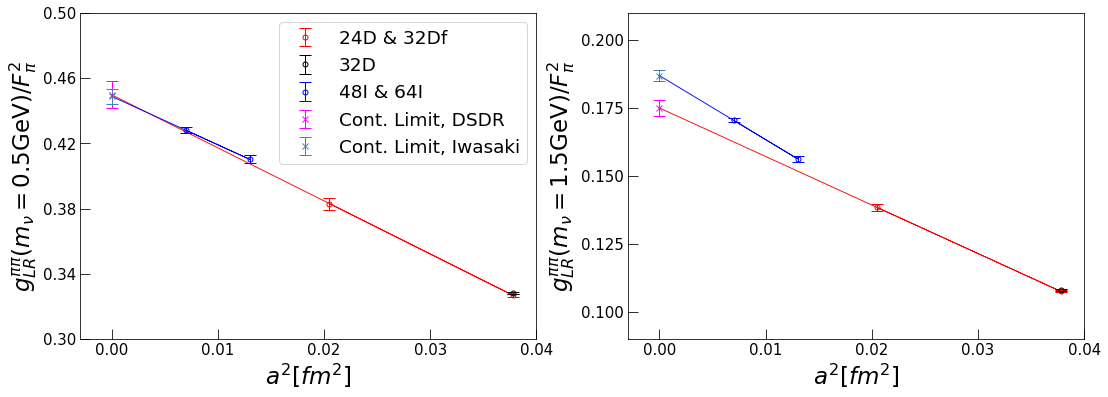}
	\caption{\label{fig:extra2}The continuum extrapolations with the linear fit ansatz of $a^2$ at some typical $m_\nu$ values. The left and right panels represent the case at $m_\nu=0.5$ GeV and $m_\nu=1.5$ GeV respectively.}
\end{figure}

The lattice results of $g_{\text{LR}}^{\pi\pi}(m_\nu)$ for five ensembles are shown on the left panel of Fig.~\ref{fig:extra}.  The 24D result is completely consistent with 32D, confirming that after applying the infinite-volume reconstruction method the finite volume effects are negligible.  
In order to control the lattice discretization error, we perform the continuum extrapolation in Iwasaki and DSDR ensembles, respectively, using a linear fit ansatz of $a^2$. The extrapolated results are shown on the right panel of Fig.~\ref{fig:extra}. In the range $m_\nu<1$ GeV, the extrapolated results from Iwasaki and DSDR are consistent within statistical errors, indicating the residual lattice artifacts are negligible. The results start to disagree at $m_\nu\gtrsim1$ GeV, indicating the residual lattice artifacts are still large. For a better illustration of these two cases, in Fig.~\ref{fig:extra2} we show the continuum extrapolations at two typical $m_\nu$ values, one smaller than 1 GeV and the other larger than 1 GeV.

We use the extrapolated result from Iwasaki ensembles as the final result and quote the deviation between Iwasaki and DSDR as the systematic uncertainty.
The corresponding result of $g_{\text{LR}}^{\pi\pi}(m_\nu)$ is given by the blue band in Fig.~\ref{fig:res1}. The numerical values are listed in Appendix~\ref{sect:appendx2}.
For comparison, we also show the results using the interpolation formula (green curve) and that using the RGE running (yellow band).

\begin{figure}
	\centering
	\includegraphics[width=0.6\textwidth]{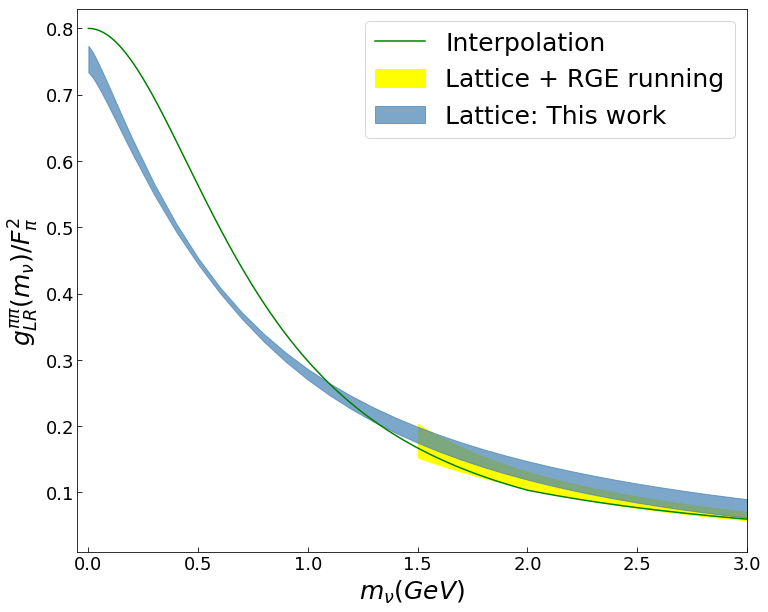}
	\caption{Comparison among the results of $g_{\text{LR}}^{\pi\pi}(m_\nu)$ using lattice QCD (blue band),
	the interpolation formula (green curve) and the RGE running of $g_4^{\pi\pi}(\mu)$ (yellow band).}
	\label{fig:res1}
\end{figure}

The idea on how to construct the interpolation formula is given in Ref.~\cite{Dekens:2020ttz}. To make Fig.~\ref{fig:res1} more comprehensive, we copy the expression of the interpolation formula here as
\begin{equation}\label{int}
	g_{\mathrm{LR}}^{\pi \pi}\left(m_{\nu}\right)=\frac{g_{\mathrm{LR}}^{\pi \pi}(0)}{1-m_{\nu}^{2} \frac{4 g_{\mathrm{LR}}^{\pi \pi}(0)}{F_{\pi}^{2}}\left[\theta\left(m_{0}-m_{\nu}\right) g_{4}^{\pi \pi}\left(m_{0}\right)+\theta\left(m_{\nu}-m_{0}\right) g_{4}^{\pi \pi}\left(m_{\nu}\right)\right]^{-1}},
\end{equation}
with $m_0=2$ GeV. The formula requires the inputs of $g_{\text{LR}}^{\pi\pi}(m_\nu)$ at $m_\nu=0$ and $g_4^{\pi\pi}(m_\nu)$ for $m_\nu\ge m_0$. For the former, it quotes the value from
Eq.~(\ref{eq:mass_splitting}). For the latter, it uses the RGE running given in Eq.~(\ref{run2}). 
Although the form seems simple, the interpolation formula captures the main $m_\nu$ dependence of $g_{\text{LR}}^{\pi\pi}(m_\nu)$, which is reasonably consistent with the lattice results up to a 20\% deviation at small $m_\nu$. Some reasons for the difference between the lattice results and the interpolating ones are explained as follows.
\begin{enumerate}
	\item At $m_\nu=0$, the interpolation formula uses the input of pion mass splitting to determine $g_{\text{LR}}^{\pi\pi}(0)$. The electromagnetic corrections to pion mass only receive the contribution of the vector current insertions, while in the lattice QCD calculation both the vector and axial-vector parts are involved. Although in the leading order $\chi$PT,
	the contributions from vector and axial-vector part are equivalent, the lattice results indicate that the higher-order correction cannot be simply neglected.
	\item When $m_\nu$ approaches zero, the interpolation formula suggests an $\mathcal{O}(m_\nu^2)$ dependence of $g_{\text{LR}}^{\pi\pi}(m_\nu)$, while 
	the lattice results yield a non-zero derivative of $g_{\text{LR}}^{\pi\pi}(m_\nu)$ with respect to $m_\nu$, see Fig.~\ref{fig:res2} for details. 
	 The reason for such behavior is explained in the Appendix~\ref{sect:appendx}.
\end{enumerate}

\begin{figure}
	\centering
	\includegraphics[width=0.6\textwidth]{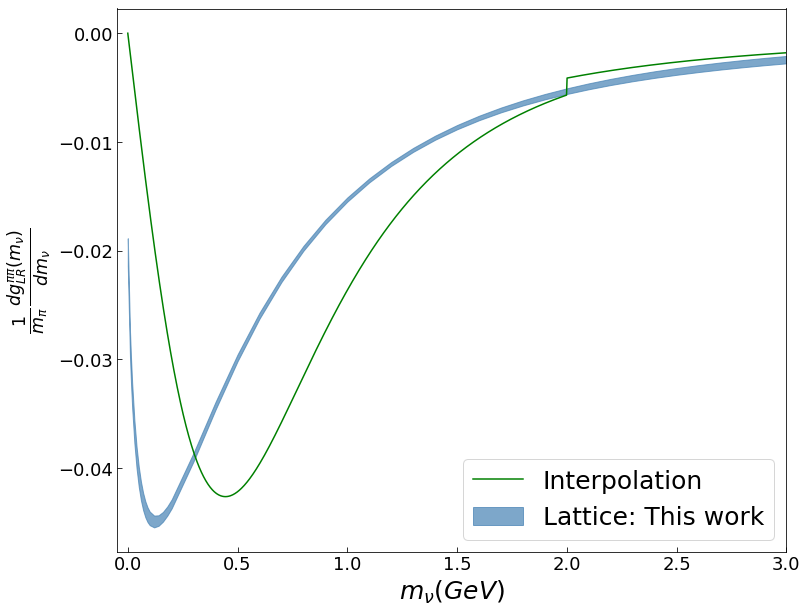}
	\caption{The derivative of $g_{\text{LR}}^{\pi\pi}(m_\nu)$ with respect to $m_\nu$. At $m_\nu=0$, the interpolation formula yields $\frac{1}{m_\pi}\frac{\mathrm{d}g_{\text{LR}}^{\pi\pi}(m_\nu)}{\mathrm{d}m_\nu}=0$ while the lattice result gives $\frac{1}{m_\pi}\frac{\mathrm{d}g_{\text{LR}}^{\pi\pi}(m_\nu)}{\mathrm{d}m_\nu}=-\frac{1}{16\pi}$. This is one of the main differences between lattice and interpolating results.}
	\label{fig:res2}
\end{figure}

At $m_\nu\gtrsim 1.5$ GeV, the lattice results of $g_{\text{LR}}^{\pi\pi}(m_\nu)$ can be compared with 
the RGE running of $g_4^{\pi\pi}(\mu)$ by setting $\mu=m_\nu$ and assuming $g_{\text{LR}}^{\pi\pi}(m_\nu)\simeq -\frac{F_\pi^2}{4m_\nu^2}g_4^{\pi\pi}(m_\nu)$.
As $g_{4,5}^{\pi\pi}(\mu)$ at $\mu=3$ GeV are determined using the lattice QCD input of pion local matrix element involving $O_4$ and $O_5$~\cite{Nicholson:2018mwc}, this comparison can be considered as a consistency check between two independent lattice QCD calculations of the bilocal matrix elements and local matrix elements. A good consistency is found in the range of 1.5 GeV $\le m_\nu\le$ 3 GeV. We should remark that the agreement is partly due to the fact that at large $m_\nu$ the errors of the current lattice results are still quite large. 
The $O(\alpha_s)$ effects neglected in the approximation of $g_{\text{LR}}^{\pi\pi}(m_\nu)\simeq -\frac{F_\pi^2}{4m_\nu^2}g_4^{\pi\pi}(m_\nu)$ may become important if the uncertainties are reduced.  

\begin{figure}
	\centering
	\includegraphics[width=0.9\textwidth]{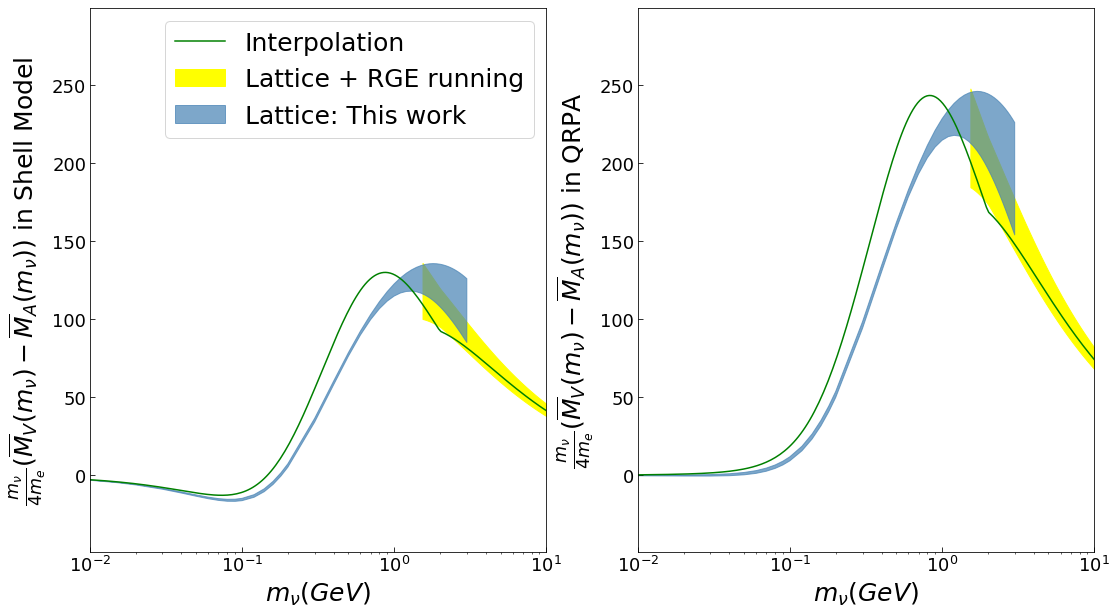}
	\caption{Comparison of the decay amplitude $m_\nu(\overline{\mathcal{M}}_V(m_\nu)-\overline{\mathcal{M}}_A(m_\nu))/(4m_e)$. Here, we take into account the impact of $g_{LR}^{\pi\pi}(m_\nu)$ and neglect $g_\nu^{NN}(m_\nu)$ and $g_{LR}^{NN}(m_\nu)$. To determine the amplitude, we use $^{136}$Xe NMEs in Ref.~\cite{Dekens:2020ttz} which are evaluated using the Shell Model~\cite{Menendez:2017fdf} (Left) and QRPA~\cite{Jokiniemi:2018sxd} (Right). The inputs of the LECs $g_{\text{LR}}^{\pi\pi}(m_\nu)$ are from lattice (blue band), interpolation formula (green curve) and RGE running (yellow band). The peak location of the decay amplitude shifts towards the larger value of $m_\nu$ when using the lattice inputs.}
\label{fig:A}
\end{figure}

We then evaluate how $g_{LR}^{\pi\pi}(m_\nu)$ affects the decay amplitude $\mathcal{A}_L$ in Eq.~(\ref{AL}). The decay amplitude is consisted of two parts, $\overline{\mathcal{M}}_V(m_\nu)-\overline{\mathcal{M}}_A(m_\nu)$ and $\overline{\mathcal{M}}_V(m_\nu)+\overline{\mathcal{M}}_A(m_\nu)$. Only the former piece receives the contributions from $g_{\text{LR}}^{\pi\pi}(m_\nu)$.
Using the lattice QCD results of $g_{\text{LR}}^{\pi\pi}(m_\nu)$ and the available $^{136}$Xe NMEs given in Ref.\cite{Dekens:2020ttz}, we update the decay amplitude, namely $m_\nu(\overline{\mathcal{M}}_V(m_\nu)-\overline{\mathcal{M}}_A(m_\nu))/(4m_e)$, in Fig.~\ref{fig:A}. 
The peak location of the decay amplitude slightly shifts towards the larger value of $m_\nu$ when using the lattice inputs. 

Here, the effects of $g_\nu^{NN}(m_\nu)$ and $g_{LR}^{NN}(m_\nu)$ are not included. They will also lead to a peak shape due to the suppression of decay amplitude at $m_\nu\to 0$ and $m_\nu\to\infty$ limits. To futher reduce the uncertainty of these LECs and determine the peak location, it is also necessary to calculate them from lattice QCD, which awaits future investigation. 

\section{Conclusion}
\label{sec4}
We perform a lattice QCD calculation of $g_{\text{LR}}^{\pi\pi}(m_\nu)$ and its derivative in the range of 0 GeV$\leq m_\nu\leq3$ GeV. Performing continuum extrapolation in the Iwasaki and DSDR ensembles separately, the lattice discretization effects are much suppressed. A good consistency with the previous lattice QCD calculation using local operators~\cite{Nicholson:2018mwc} is found. 
Comparing our results with the naive interpolation formula at small $m_\nu$, we find a $\sim20$ \% deviation. The main reasons for the difference are two folds. First, the use of 
the pion mass splitting given in Eq.~(\ref{eq:mass_splitting}) provides a reasonably good but not sufficiently accurate value for $g_{\text{LR}}^{\pi\pi}(0)$. Second, we obtain the non-zero derivative of  $g_{\text{LR}}^{\pi\pi}(m_\nu)$ with respect to $m_\nu$ at $m_\nu=0$ while the interpolation formula yields zero. 
Using $g_{\text{LR}}^{\pi\pi}(m_\nu)$ determined from lattice QCD, we present the neutrino mass dependence of single sterile neutrino contribution to $0\nu2\beta$ decay amplitude. 
As pointed out by the Snowmass white paper~\cite{Cirigliano:2022oqy}, the EFTs that systematically describe $0\nu2\beta$ decay amplitude in the few-nucleon and pion sector need to be complemented with values for LECs from lattice QCD.
Our work serves as one more example for such purpose.

\acknowledgements

We gratefully acknowledge many helpful discussions with our colleagues from the
RBC-UKQCD Collaborations.
We thank W. Dekens, J. de Vries, K. Fuyuto and E. Mereghetti for bringing our attention to the sterile neutrino contributions to $0\nu2\beta$ decays. 
X.F. and X.Y.T. were supported in part by NSFC of China under Grants
No. 12125501, No. 12070131001, and No. 12141501,
and National Key Research and Development Program of China under No. 2020YFA0406400.
L.C.J. acknowledges support by DOE Office of Science Early Career Award No. DE-SC0021147
and DOE Award No. DE-SC0010339. The research reported in this work was carried out using the computing facilities at Chinese National Supercomputer Center in Tianjin. It also made use of computing and long-term storage facilities of the USQCD Collaboration, which are funded by the Office of Science of the U.S. Department of Energy.

\appendix

\section{Numerical values for $g_{LR}^{\pi\pi}(m_\nu)$ on lattice\label{sect:appendx2}}
In Table~\ref{values} we list the lattice results of $g_{\text{LR}}^{\pi\pi}(m_\nu)$, which can be used in the future phenomenological studies.

\begin{table*}
	\begin{ruledtabular}
		\begin{tabular}{cc|cc|cc|cc}
			$m_\nu$ [GeV] & $g_{\text{LR}}^{\pi\pi}(m_\nu)/F_\pi^2$&$m_\nu$ [GeV] & $g_{\text{LR}}^{\pi\pi}(m_\nu)/F_\pi^2$&$m_\nu$ [GeV] & $g_{\text{LR}}^{\pi\pi}(m_\nu)/F_\pi^2$&$m_\nu$ [GeV] & $g_{\text{LR}}^{\pi\pi}(m_\nu)/F_\pi^2$\\
			\hline
			0&0.754(20)			&0.2&0.622(11)		&1.2&0.235(10)		&2.2&0.118(14)\\
			0.02&0.745(19)		&0.3&0.557(8)		&1.3&0.217(11)		&2.3&0.111(14)\\
			0.04&0.733(18)		&0.4&0.499(6)		&1.4&0.201(11)		&2.4&0.105(14)\\
			0.06&0.720(17)		&0.5&0.449(5)		&1.5&0.187(12)		&2.5&0.099(14)\\
			0.08&0.707(16)		&0.6&0.405(4)		&1.6&0.174(13)		&2.6&0.094(14)\\
			0.10&0.693(15)		&0.7&0.366(5)		&1.7&0.162(13)		&2.7&0.089(14)\\
			0.12&0.679(14)		&0.8&0.333(6)		&1.8&0.152(13)		&2.8&0.084(14)\\
			0.14&0.664(13)		&0.9&0.304(7)		&1.9&0.142(14)		&2.9&0.080(14)\\
			0.16&0.650(13)		&1.0&0.278(8)		&2.0&0.133(14)		&3.0&0.076(14)\\
			0.18&0.636(12)		&1.1&0.255(9)		&2.1&0.125(14)		& & \\
		\end{tabular}
	\end{ruledtabular}
	\caption{Lattice results of $g_{\text{LR}}^{\pi\pi}(m_\nu)$ with $m_\nu$ from 0 GeV to 3 GeV.}
	\label{values}
\end{table*}

\section{\boldmath$g_{\text{LR}}^{\pi\pi}(m_\nu)$ at small neutrino mass}
\label{sect:appendx}

As scalar propagator $S_0^E(x)$ contains a mass-square term, it seems that $g_{\text{LR}}^{\pi\pi}(m_\nu)$ shall have a linear dependence on $m_\nu^2$ when $m_\nu\to0$. However, we will demonstrate here that this is not the case.
We write down the pion intermediate-state contribution to the hadronic function $H_{VV}^E(x)$ as
\begin{equation}
H_\pi(x)=-\int \frac{d^3 p}{(2\pi)^3 2E_{\pi,p}}m_\pi(m_\pi+E_{\pi,p})[F^{(\pi)}(q^2)]^2 e^{-(E_{\pi,p}-m_\pi)|t|}e^{i\vec{p}\cdot\vec{x}}
\end{equation}
where $F^{(\pi)}(q^2)$ is pion electromagnetic form factor introduced by using the isospin rotation. $q^2=2m_\pi(m_\pi-E_{\pi,p})$ comes from the momentum transfer between the charged pion in the initial state and the neutral pion in the intermediate state. The contribution of $H_\pi(x)$ propagates into $g_{\text{LR}}^{\pi\pi}(m_\nu)$ and yields
\begin{equation}\label{Gpi}
g_{\text{LR},\pi}^{\pi\pi}(m_\nu)=\frac 18\int\frac{d^3p}{(2\pi)^3}\frac{m_\pi(m_\pi+E_{\pi,p})[F^{(\pi)}(q^2)]^2}{E_{\pi,p}E_{\nu,p}(E_{\pi,p}-m_\pi+E_{\nu,p})},
\end{equation}
where we define $E_{\pi,p}^2=p^2+m_\pi^2$ and $E_{\nu,p}^2=p^2+m_\nu^2$.

At small $m_\nu$, we can take a scale $\Lambda$ satisfying $m_\nu \ll \Lambda \ll m_\pi$.
In the momentum region $p<\Lambda$, the integral can be simplified as
\begin{equation}
g_{\text{LR},\pi}^{\pi\pi}(m_\nu)|_{p<\Lambda}=\frac{m_\pi}{8\pi^2}\int_0^\Lambda \mathrm{d}p\, \frac{p^2}{p^2+m_\nu^2}\left(1+\mathcal{O}\left(\frac{p^2}{m_\pi^2}\right)\right).
\end{equation}
We then consider the derivative of
\begin{equation}
\begin{aligned}
\frac{\mathrm{d}g_{\text{LR},\pi}^{\pi\pi}(m_\nu)|_{p<\Lambda}}{\mathrm{d}m_\nu}&=-\frac{m_\pi m_\nu}{4\pi^2}\int_0^\Lambda\mathrm{d}p\, \frac{p^2}{(p^2+m_\nu^2)^2}\left(1+\mathcal{O}\left(\frac{p^2}{m_\pi^2}\right)\right)\\
&=-\frac{m_\pi m_\nu}{4\pi^2} \left(\frac{\pi}{4m_\nu}-\frac{1}{2\Lambda}+\cdots\right).
\end{aligned}
\end{equation}
In the momentum region of $p\ge \Lambda$, one can simply take a Taylor expansion of $1/(p^2+m_\nu^2)=1/p^2-m_\nu^2/p^4+\cdots$. The derivative yields zero at $m_\nu=0$. Thus, we can conclude that 
\begin{equation}
\label{dg0}
\frac{\mathrm{d}g_{\text{LR},\pi}^{\pi\pi}(m_\nu)}{\mathrm{d}m_\nu}\Big|_{m_\nu= 0}=-\frac{m_\pi}{16\pi}.
\end{equation}
For the contributions from the excited states, as the integral only contains a factor of $1/\sqrt{p^2+m_\nu^2}$ rather than $1/(p^2+m_\nu^2)$ at small $m_\nu$, one can show that the derivative vanish when $m_\nu\to0$.
This means that $\frac{\mathrm{d}g_{LR}^{\pi\pi}(m_\nu)}{\mathrm{d}m_\nu}\big|_{m_\nu= 0}$ is a structure-independent value. In Fig.~\ref{fig:res2}, we obtain the lattice result of $\frac{1}{m_\pi}\frac{\mathrm{d}g_{LR}^{\pi\pi}(m_\nu)}{\mathrm{d}m_\nu}\big|_{m_\nu= 0}=-0.0201(16)$, which is well consistent with the expectation of $-1/(16\pi)\approx -0.0199$.

\bibliography{ref}

\end{document}